\documentclass[12pt]{elsart}
\usepackage{psfig}
\usepackage{subfigure}
\makeatletter
\newcount\@tempcntc
\def\@citex[#1]#2{\if@filesw\immediate\write\@auxout{\string\citation{#2}}\fi
  \@tempcnta\z@\@tempcntb\m@ne\def\@citea{}\@cite{\@for\@citeb:=#2\do
    {\@ifundefined
       {b@\@citeb}{\@citeo\@tempcntb\m@ne\@citea\def\@citea{,}{\bf ?}\@warning
       {Citation `\@citeb' on page \thepage \space undefined}}%
    {\setbox\z@\hbox{\global\@tempcntc0\csname b@\@citeb\endcsname\relax}%
     \ifnum\@tempcntc=\z@ \@citeo\@tempcntb\m@ne
       \@citea\def\@citea{,}\hbox{\csname b@\@citeb\endcsname}%
     \else
      \advance\@tempcntb\@ne
      \ifnum\@tempcntb=\@tempcntc
      \else\advance\@tempcntb\m@ne\@citeo
      \@tempcnta\@tempcntc\@tempcntb\@tempcntc\fi\fi}}\@citeo}{#1}}
\def\@citeo{\ifnum\@tempcnta>\@tempcntb\else\@citea\def\@citea{,}%
  \ifnum\@tempcnta=\@tempcntb\the\@tempcnta\else
   {\advance\@tempcnta\@ne\ifnum\@tempcnta=\@tempcntb \else \def\@citea{--}\fi
    \advance\@tempcnta\m@ne\the\@tempcnta\@citea\the\@tempcntb}\fi\fi}
\makeatother
\begin{document}
\begin{frontmatter}
\title{Electromagnetic Polarizabilities of Nucleons bound in
$^{40}$Ca, $^{16}$O and $^4$He \thanksref{dfg}}
\thanks[dfg]{Supported by Deutsche Forschungsgemeinschaft (contracts
Schu 222 and DFG 438-113-173), the Deutscher Akademischer Austauschdienst,
the Swedish Natural Science Research Council, the Knut and Alice Wallenberg 
Foundation, the Crafoord Foundation and the Swedish Institute.}
\author[goe]{S. Proff},
\author[goe]{C. P\"och},
\author[goe]{T. Glebe},
\author[lu]{J.-O. Adler}, 
\author[lu]{K. Fissum},
\author[lu]{K. Hansen}, 
\author[goe]{M.-Th. H\"utt}, 
\author[goe]{O. Kaltschmidt},  
\author[lu]{M. Lundin}, 
\author[lu]{B. Nilsson},       
\author[lu]{B. Schr\"oder}, 
\author[goe]{M.~Schumacher}\footnote{e-mail: schumacher@physik2.uni-goettingen.de},
\author[lu]{D. Sims},
\author[goe]{F. Smend}, 
\author[goe]{F. Wissmann}
\address[goe]{Zweites Physikalisches Institut, 
             Universit\"at G\"ottingen, D-37073
             G\"ottingen, Germany}
\address[lu]{Department of Physics, University of Lund, S-22362 Lund, Sweden}
\begin{abstract}
Differential cross sections for elastic scattering of photons
have been measured for $^{40}$Ca at energies of 58 and 74 MeV and for 
$^{16}$O and $^4$He at 61 MeV, in the angular range from 45$^{\circ}$ 
to 150$^{\circ}$. Evidence is obtained
that there are no significant in-medium modifications of the electromagnetic 
polarizabilities except for those originating from meson exchange currents.
\end{abstract}
\end{frontmatter}
\noindent
$Keywords:$ Measured: Compton scattering by $^{40}$Ca, $^{16}$O
and $^4$He. Extracted: In-medium
electromagnetic polarizabilities of the nucleon.\\ 
\section{Introduction}
%
%
The electromagnetic polarizabilities belong to the few fundamental
structure constants of the nucleon which  can be measured in the 
free and bound state with comparatively high precision. These quantities, 
therefore, belong to those  observables where possible in-medium 
modifications of nucleon properties  may manifest themselves.

Though electromagnetic polarizabilities of hadrons have been discussed 
already for a long time, some clarifying remarks concerning these quantities
are advisable at the beginning. The electromagnetic
polarizabilities ${\bar \alpha}_p$ and ${\bar \beta}_p$ of the free proton
defined through the Baldin-Lapidus sum rule or through the differential 
cross section for Compton scattering of real photons (RCS) correspond to the
only possible complete (gauge invariant)  definitions at the photon point
\cite{lvov93}.
Commonly used expressions like 
${\bar \alpha} = { \alpha_{\circ} + \Delta \alpha}$ and
${\bar \beta} = \beta_{para} + \Delta \beta$ lead to model dependent 
quantities
on the r.h.s. of the two equations   which cannot be measured by any 
experiment. For the neutron an additional quantity  $\alpha_n$ is defined, 
denoting the 
electric polarizability determined through Coulomb scattering. This
quantity is slightly different from the gauge invariant
quantity ${\bar \alpha}_n$ with  ${\bar \alpha}_n - \alpha_n = 0.62$
(in units of $10^{-4} fm^3$)  \cite{lvov93}. For a clear-cut definition 
of the in-medium polarizabilities of the nucleon \cite{rosa85,schumacher88}
we have to distinguish between (i)  
effects of the internal structure of the nucleon 
entering into ${\tilde \alpha}_N$ and ${\tilde \beta}_N$ 
which are  one-body quantities and may be considered as the true
in-medium electromagnetic polarizabilities,  and (ii) two-body effects of meson
exchange currents between p-n pairs which may be denoted by 
${\delta \alpha}$ and ${\delta \beta}$. From a phenomenological
point of view these latter  quantities are not part of the in-medium
electromagnetic polarizabilities but are used to parametrize 
that part of the ``mesonic seagull amplitude'' 
(Thomson scattering by correlated p-n pairs)
which is energy dependent in the forward direction.  (For more
details see section  3.3).
In a first step of the  analysis of experimental data the convolution
of the true in-medium electromagnetic polarizabilities and the meson 
exchange corrections are  determined. This leads to the effective
in-medium  electromagnetic polarizbilities 
${\tilde \alpha}_{eff} = {\tilde \alpha}_N + {\delta \alpha}$ and
${\tilde \beta}_{eff} = {\tilde \beta}_N + {\delta \beta}$.
The quantities $\delta \alpha$ and $\delta \beta$ have to be 
calculated and, therefore, introduce some model dependence into the
determination of ${\tilde \alpha}_N $ and ${\tilde \beta}_N$. 
However, as will be 
shown later, the respective model dependent uncertainties are not large.

In  nuclei with $Z=N$ which are investigated here
the arithmetic averages 
of the proton and neutron electromagnetic polarizabilities are observed.
From the Baldin-Lapidus sum rule
applied to experimental photoabsorption cross sections it is 
known \cite{lvov79,babusci98} that the sum ${\bar \alpha}_N + 
{\bar \beta}_N = 15.0 \pm 0.4$ (in units of
$10^{-4}\mbox{fm}^3$ with $\bar{\alpha}_N=11.3\pm1.5$,
$\bar{\beta}_N=3.7\mp1.5$  
\cite{lvov93,PDG96}) of 
free-nucleon polarizabilities is only 
slightly shifted down in the nuclear medium to about 
${\tilde \alpha}_{eff} +\tilde{\beta}_{eff}=14.0$ \cite{ludwig92}. 
This observation of an approximately constant sum of effective 
electromagnetic polarizabilities does not
exclude that the relative sizes of the effective in-medium electric ${\tilde 
\alpha}_{eff}$
and magnetic $\tilde{\beta}_{eff}$ polarizabilities 
may be considerably different
from the corresponding free-nucleon values due to meson exchange-currents
and/or modifications of the  internal structure of the  nucleon. 

In this connection 
it is of great interest that recently there was the claim \cite{feldman96} 
that in-medium shifts as large as $\Delta{\alpha}=- 8$ and $\Delta{\beta}=+8$ 
compared to the free-nucleon electromagnetic polarizabilities
have been observed in Compton scattering experiments on $^{16}$O.
In-medium shifts of this large size would imply that the dominance of the
electric multipolarity observed for the free-nucleon electromagnetic 
polarizabilities is replaced  by a dominance of the magnetic 
multipolarity for the case of bound nucleons. This shift of multipolarity
$-$ if confirmed  $-$ has the potentiality of being an important 
discovery, unless it can be  traced back to meson exchange 
currents as tentatively assumed in 
\cite{feldman96}. However, according to our present knowledge the effects
of meson exchange currents are not large enough to explain the 
multipolarity shifts reported in \cite{feldman96}. 
Levchuk \cite{levchuk97}  has carried out an explicit calculation of 
the amplitudes for Compton
scattering by $^2$H. This  calculation includes all relevant 
effects and, hence, also the modification of the electromagnetic
polarizabilities due to meson exchange-currents. His results are 
\cite{levchuk97} ${\delta \alpha}(^2\mbox{H})= - 0.9$ and 
${\delta \beta}(^2\mbox{H}) \approx 0$ which apparently are much too small
to take care of the large in-medium modifications observed in 
\cite{feldman96}. These conclusions concerning the effects of meson
exchange currents on the in-medium electromagnetic polarizabilities 
have been confirmed in two recent 
theoretical investigations \cite{huett96,huett98} carried out for 
complex nuclei which will be discussed in detail in  section 3.

Unfortunately, the findings of \cite{feldman96} go along with 
a discrepancy between 
two sets of experimental data obtained for Compton scattering by $^{16}$O
carried out in the quasideuteron range \cite{feldman96,haeger95}.
In the present paper we, therefore, concentrate on collecting new 
experimental data which are of relevance for the investigation of
in-medium electromagnetic polarizabilities. Three different nuclei have
been investigated, viz. $^{40}$Ca, $^{16}$O and $^4$He for the following 
reasons:\\ 
(i) All three nuclei have carefully been studied by photoabsorption
experiments   \cite{ahrens75,arkatov78}, so that the 
resonance amplitude
$R(E,\theta)$ describing the nuclear-structure dependent part of the
Compton scattering process is known with fairly good precision. This
resonance amplitude $R(E,\theta)$  has the character of a background 
underneath the amplitude $T_N$ for Compton scattering through virtual
excitation of the internal degrees of freedom of the nucleon, which is the 
scattering processes we are interested in here. Close to the forward
direction the amplitude $T_N$ is given by the well known sum of effective
in-medium electromagnetic polarizabilities 
${\tilde \alpha}_{eff} + {\tilde \beta}_{eff}$
so that the predictions obtained for $R(E,\theta)$ can be tested, and 
adjusted to fit the small-angle data where necessary. These adjustments
are legitimate as long as they remain within the experimental errors of 
the photoabsorption cross sections.
The only free
parameter then is the quantity 
${\tilde \alpha}_{eff} - {\tilde \beta}_{eff}$ which can 
be extracted from the angular dependence of the differential cross section. \\
(ii) The calculated meson-exchange corrections  of the
in-medium electromagnetic polarizabilities show an
$A$-dependence \cite{huett98}. Therefore, experiments at 
different mass number $A$ 
are well motivated.\\
(iii) For $^{40}$Ca Compton scattering experiments have been carried out 
in the giant-resonance energy-range \cite{wright85} 
giving additional information on the 
nuclear-structure part $R(E,\theta)$ of the Compton scattering process.
This also helps to improve on the reliability of our conclusions concerning 
the amplitude $R(E,\theta)$ in the quasi-deuteron energy-range where the 
present experiments are carried out. Furthermore, for this nucleus 
there is only a very
small A dependence of the meson exchange corrections 
of the in-medium
electromagnetic polarizabilities $\delta \alpha$ and $\delta \beta$, 
making the predictions very reliable.
Therefore, $^{40}$Ca was used in a first place to search for indications
of the  large modifications $\Delta \alpha = -$8, $\Delta \beta = +$8
suggested in  \cite{feldman96}. \\
(iv) The nucleus $^{16}$O has been selected because this is the one 
where a discrepancy 
between our previous experimental  data \cite{haeger95} and the 
Illinois/Saskatoon data  \cite {feldman96} 
was observed. Furthermore, at this nucleus there is a sizable finite-size
effect of the quantities $\delta \alpha$ and $\delta \beta$ which may be
tested by comparison with experimental differential cross sections.\\
(v) The  nucleus $^4$He has the advantage of combining the  
structure of a few-nucleon system with the large binding energy 
of a complex nucleus. These favourable properties and the fact that
$-$ due to its small size $-$
formfactors have only a comparatively small effect on the large-angle
differential cross section, make it especially suited for studies of
in-medium electromagnetic polarizabilities. Irrespective of this, 
there were no
reliable data for this nucleus from which conclusions on the in-medium
electromagnetic polarizabilities could be drawn. The data published in 
\cite{wells92} are restricted to the energy range from 20 MeV to 35 MeV 
where the sensitivity to the electromagnetic polarizabilities is small.
The angular dependence of differential cross sections measured by us
in an early experiment at 87 MeV
\cite{fuhrberg95}
in principle is very suitable for extracting  in-medium electromagnetic
polarizabilities. However, these data were obtained using untagged
bremsstrahlung and, therefore, $-$ though carried out very carefully $-$
may have suffered from the systematic uncertainties inherent in that method.
When using untagged bremsstrahlung the determinations of the 
numbers of primary and secondary photons are carried out through arbitrary
cuts in continuous spectra. The shortcomings of this method have been
eliminated as far as possible by carrying out Monte Carlo simulations.
Nevertheless, 
from our present point of view we have good 
reasons  to distrust the bremsstrahlung method and 
to disregards these previous data \cite{fuhrberg95}.
A further set of data on $^4$He extending up to
73 MeV contained in a thesis \cite{wells90}  has not been published up to 
now. Since these data have been measured using tagged photons  they are
on an equal footing with our present data and, therefore, should 
be shown  together with them.
\section{Experiment}
The experiment has been carried out at the high-resolution tagging spectrometer
of the MAX-laboratory in Lund (Sweden) \cite{adler97}. Making use of
the high duty-factor 95 MeV electron beam of the MAX stretcher ring, this 
tagging spectrometer was operated in two different settings to provide 
photons in the energy ranges around 60 MeV and  74 MeV. The average count 
rate per tagger channel was 450 kHz, thus keeping accidental stops of 
correlated events (stolen trues) lower than 6 \%. This rate was 
calculated and eliminated by a correction. For one recent experiment on
$^{16}$O we increased the count-rate per tagger channel by a factor
of 3 $-$ 4 , in order to improve on the statistical precision. These data
will be discussed separately in section 4.

The NaI(Tl)-detector set-up described in detail in \cite{haeger95}
has been used in the present experiment to  measure 
differential cross sections. Four detectors are available to be operated 
simultaneously at four or three different scattering angles.
The diameters of these NaI(Tl)-detectors were 25.4 cm, the lengths 25.4 cm 
for three
of them and 35.6 cm for one of them. Each of the detectors was surrounded 
by shields of plastic scintillators and lead. The plastic scintillator
shields surrounded the detectors on four sides. They were operated as 
anti-cosmic shields only, by setting the thresholds high enough to minimize 
rejection of photon events where part of the shower had escaped into
the plastic scintillators. In order to correct for possible residual
losses, a full GEANT-code simulation of the detection process has been 
carried out. In addition, the rates of scattered photons and direct 
photons, both normalized to the rates of electrons in the tagger channels, 
have been detected under identical conditions, thus largely eliminating
threshold effects in the ratio of the two quantities.

The lead shields had 
cylindrical openings at the front sides serving as collimators for the 
incident photons. 
These collimator openings were covered by  additional 5mm thick plastic 
scintillators, to detect charged particles accompanying  the beam of 
scattered photons. The widths of the collimators were chosen such, that 
each part of the incoming photon beam was bound to pass the total depth of 
the NaI(Tl) detector if not absorbed in the detector volume. 

In order to suppress neutron backgrounds these 
collimator openings were shielded by material containing $^6$Li or
$^9$Be. For part of the detectors $^6$Li shields were also mounted 
between the plastic scintillators and the NaI(Tl) detectors. Residual
count-rates due to neutrons were safely eliminated through cuts in the
spectra of time differences between signals from the NaI(Tl) detectors
and the correlated signals from the tagger channels.  
  
The quality of the spectra of scattered photons and the separation of 
Compton events from background events has been documented in
our previous publications \cite{ludwig92,haeger95,fuhrberg92}.
\section{Discussion of Theoretical Results}
\subsection{Status of the free polarizabilities of the nucleon}
The numbers given in the introduction for the free-nucleon electromagnetic 
polarizabilities are based on information collected in the Review of Particle 
Physics 1996 \cite{PDG96}. Meanwhile,
some additional work has been carried out which has to be discussed here.
Experiments on Compton scattering by the proton in the $\Delta$ resonance
range carried out at MAMI (Mainz)  and other data led to the
conclusion \cite{huenger97} that the $M_{1+}$ amplitude of the
SAID(SM95) parametrization of photo meson amplitudes \cite{arndt96} had to 
be reduced by about 2.8\%. This finding initiated a  re-evaluation of the
parametrization of photo-meson amplitudes now available as SAID(SP97K)
which to a large extent confirmed the findings of \cite{huenger97}.
On the basis of this new parametrization, SAID(SP97K), Babusci et al.
\cite{babusci98} carried out a new evaluation of the Baldin-Lapidus
sum rule and obtained ${\bar \alpha}_p +{\bar \beta}_p=13.69 \pm 0.14$
(in units of $10^{-4} fm^3$)
instead of the previous ${\bar \alpha}_p +{\bar \beta}_p=14.2 \pm 0.3$
for the proton and ${\bar \alpha}_n +{\bar \beta}_n=14.40 \pm 0.66$
instead of the previous ${\bar \alpha}_n +{\bar \beta}_n=15.8 \pm 0.5$
for the neutron. Hence, the new proton-neutron average would be
${\bar \alpha}_N +{\bar \beta}_N=14.0 \pm 0.4$ instead of
${\bar \alpha}_N +{\bar \beta}_N=15.0 \pm 0.4$.
Though, apparently there is some room for discussion,
the overall precision of ${\bar \alpha}_N +{\bar \beta}_N$ is by far
good enough for the purpose of our present data analysis. Furthermore,
adjustments in the predicted differential cross sections are possible at 
small angles as discussed in section 1, making small differences
in ${\bar \alpha}_N +{\bar \beta}_N$ unobservable. Note that the free 
parameter of our data analysis is the difference of in-medium
polarizabilities  ${\tilde \alpha}_N - {\tilde \beta}_N$ which 
$-$ within limits $-$ has no influence on the differential cross 
section at zero angle.
%

Another word of precaution has to be devoted to the multipole 
decomposition of the free-nucleon electromagnetic polarizabilities.
The electric polarizability ${\bar \alpha}_p$ of the proton given by the 
(PDG96) \cite{PDG96} is based on the ``global average'' of MacGibbon et al.
\cite{macgibbon95} taking into account the Baldin-Lapidus sum rule
according to the evaluation of \cite{lvov79}
and  experimental data on Compton scattering by the proton. 
The first meaningful number for the electric polarizability of the neutron 
was measured by Rose et al. \cite{rose90}
through quasifree Compton scattering on neutrons
bound in the deuteron. The precision of this experiment was surpassed
by an experiment on scattering of neutrons in the  Coulomb field of Pb
nuclei, enriched in $^{208}$Pb \cite{schmied91}. At this stage 
of development it was believed that with ${\bar \alpha}_p =12.1 \pm 0.8 
(\mbox{stat}) \pm 0.5 (\mbox{syst})$ \cite{macgibbon95} and  
${\bar \alpha}_n =12.6 \pm 1.5 (\mbox{stat}) \pm 2.0 
(\mbox{syst})$ \cite{schmied91} the electric polarizabilities of the two
nucleons were equal to each other within rather good levels of
precision. The number 12.6 given for $\bar \alpha_n$  contains a small
correction of 0.62 which takes care of the fact that the quantity 
$\alpha_n$ extracted from Coulomb scattering experiments is slightly 
smaller than the electric polarizability $\bar \alpha_n$ \cite{lvov93}. 
Meanwhile the situation concerning the electric polarizability
of the neutron has changed due to the 
fact that there is a new experiment on Coulomb scattering on the neutron
\cite{koester95}
leading to an electric polarizability compatible with $\alpha_n =0$ and 
due to the fact that the error of the former Coulomb scattering result  
\cite{schmied91}
may possibly have been  underestimated \cite{enik97}. A detailed  discussion
of this point and a possible experimental way out of this problem has recently 
been discussed by Wissmann et al. \cite{wissmann98}.

For the present data analysis the following way of arriving at free-nucleon
electromagnetic polarizabilities appears to be appropriate:\\
(i) For the protons we use the global average ${\bar \alpha}_p = 12.1 \pm 
0.9 $.\\
(ii) For the neutron we use the average given by PDG96 \cite{PDG96}, 
which amounts to 
${\bar \alpha}_n = 10.4 \pm 2.1$ after including the difference between 
${\bar \alpha}_n$ and $\alpha_n$. This number is very close to the 
Compton scattering result of Rose et al. \cite{rose90} and will not change
too much in case the error of Schmiedmayer's Coulomb scattering result 
\cite{schmied91}
would be larger than anticipated.\\
(iii) For the free nucleon we use the weighted average of the two numbers
given in (i) and (ii)
with the weights given by the numbers of nucleons. As a result we arrive at
${\bar \alpha}_N = 11.3 \pm 1.5$ and ${\bar \beta}_N=3.7 \mp 1.5$ as 
given in the introduction. 
\subsection{Scattering amplitudes}
For purpose of  determining  in-medium electromagnetic polarizabilities it
is convenient to write the total Compton amplitude of the nucleus 
$T_{tot}(E,\theta)$
as a  superposition of single-nucleon amplitudes and a nuclear contribution,
$viz.$
\begin{equation}
\label{amplitudeTA}
T_{tot}(E,\theta)= \left[ Z T_p(E,\theta)  + N T_n(E,\theta)  \right] F_1(q) 
+ S(E,\theta) + R(E,\theta).
\end{equation}
In (\ref{amplitudeTA}) $Z$ and $N$ are the proton and neutron numbers, 
respectively, and $F_1(q)$ the one-body formfactor describing
the distribution of nucleons in the nucleus.
The amplitudes $T_p(E,\theta)$ and $T_n(E,\theta)$ for single protons and
single neutrons, respectively, differ from the corresponding 
free-nucleon amplitudes
by the fact that spin dependent terms cancel in a spin-saturated nucleus. For 
energies below meson photoproduction threshold  these amplitudes may be 
written in the from
%
%
\begin{eqnarray}
\label{amplitudesTN}
T_p(E,\theta) &=& -\frac{e^2}{Mc^2} + \tilde \alpha_{p} \,
\left(\frac{E}{\hbar c}\right)^{2} \, g_{E1}(\theta) + \tilde \beta_{p} \,
\left(\frac{E}{\hbar c}\right)^{2} \, g_{M1}(\theta) +  O(E^{4})
 \nonumber\\
T_n(E,\theta) &=& \hspace{16mm} \tilde \alpha_{n} \,
\left(\frac{E}{\hbar c}\right)^2 \, g_{E1}(\theta) + \tilde \beta_{n} \,
\left(\frac{E}{\hbar c}\right)^2 \, g_{M1}(\theta) +  O(E^4).
\end{eqnarray}
When inserted into (\ref{amplitudeTA}), the Thomson term of 
(\ref{amplitudesTN}) leads to the  kinetic seagull amplitude
$B(E,\theta)$, whereas the other terms lead to the amplitude
\begin{equation}
T_N(E,\theta)=A\left( {\tilde \alpha}_N g_{E1}(\theta)
+ {\tilde \beta}_N g_{M1}(\theta) \right) {\left( \frac{E}{\hbar c} 
\right)}^2 F_1(q) +(E^4),
\label{NuclAmpl}
\end{equation}
with
\begin{equation}
{\tilde \alpha}_N =\frac{Z}{A}{\tilde \alpha}_p + \frac{N}{A}{\tilde 
\alpha}_n, \hspace{15mm} {\tilde \beta}_N =\frac{Z}{A}{\tilde \beta}_p + 
\frac{N}{A}{\tilde \beta}_n.
\label{alpha-beta}
\end{equation} 
In (\ref{amplitudesTN}) and (\ref{NuclAmpl}) 
the quantities $g_{E1}(\theta)$ and  
$g_{M1}(\theta)$ are the angular distribution functions for  
electric and magnetic dipole scattering, respectively.
The free-nucleon electromagnetic polarizabilities
${\bar \alpha}$ and ${\bar \beta}$ have been replaced by the corresponding
bound-nucleon quantities, ${\tilde \alpha}$ and ${\tilde \beta}$, because 
of a possible change of these quantities due to a modification of the 
internal nucleon structure. Theoretical discussions may be found in  
\cite{schumacher88,drechsel84,lvov92}.
The spin-independent parts of the Born terms are represented
by the Thomson amplitude only, though there are additional terms which
depend on the anomalous magnetic moments. These terms have been discussed 
in \cite{haeger95}. They proved to be very small, corresponding to a 
modification of the electromagnetic polarizabilities of the order 
$0.3\times10^{-4}\mbox{fm}^3$. Because of this smallness and because of
some uncertainties in including them into the nuclear amplitudes we 
decided to neglect them here. The resonance (nuclear structure)
amplitude $R(E,\theta)$ and the mesonic seagull amplitude $S(E,\theta)$ 
have also been discussed in \cite{haeger95}. For the present purpose
it is enough to say that $R(E,\theta)$ can be obtained from the nuclear
photoabsorption cross section via the optical theorem and the once-subtracted
dispersion relation. For practical purposes it is convenient to represent
the nuclear photoabsorption cross section by a superposition of 
Lorentzian lines. In the quasi-deuteron energy-range the amplitude
$R(E,\theta)$  is mainly determined by the real parts of the
giant-resonance amplitudes. The quasi-deuteron contribution is small 
because its real part has a zero crossing in this range and its imaginary part 
is suppressed because there is no partner to interfere with. 
The main component of the mesonic seagull amplitude $S(E,\theta)$,
i.e. its ``static'' approximation
${\tilde S}(E,\theta)$, is completely fixed
through the enhancement constant $\kappa$ of the nuclear photoabsorption
cross section and the two-body formfactor $F_2(q)$, describing the 
distribution of correlated proton-neutron pairs in the nucleus. 
Up to recently,  the two-body formfactor was customarily calculated
using the relation $F_2(q) = \left( F_1(q/2) \right)^2$ which is valid 
under the assumption that nuclear interaction through meson exchange is 
of infinite range. A more refined calculation with a realistic 
nucleon-nucleon interaction has been carried out in \cite{huett98}.
The result showed that $F_2(q)$  for $A = 40$  may be reproduced 
by  $F_1(q)$ for $A=22$. We use this latter prediction,
though for our present purpose the use of the one or 
the other representation of  $F_2(q)$ makes no essential difference.
This is even more the case for the smaller nucleus $^{16}$O.
Nevertheless, we applied the exact result of the model calculation 
\cite{huett98}  also for for $^{16}$O.    For 
$^4$He the relation $F_2(q) = \left( F_1(q/2) \right)^2$ becomes
exact.
Deviations of the mesonic seagull amplitude    $S(E,\theta)$ from its
static approximation will be discussed in the next subsection.
\subsection{Meson exchange currents}  
In complex nuclei meson exchange currents are a consequence of the
interaction between proton-neutron
pairs (quasideuterons). The main part of this effect is contained 
in the ``static'' mesonic seagull amplitude ${\tilde S}(E,\theta)$.
Deviations from this approximation
 may be written in the form
\begin{eqnarray}
\label{DeltaM}
S(E,\theta) - {\tilde S}(E,\theta) 
& = &\left(\delta \alpha \, g_{E1}(\theta) + 
\delta \beta \, g_{M1}(\theta)  \right) \left( \frac{E}{\hbar c} 
\right)^2 F_2(q) \nonumber \\
&\approx& \left(\delta \alpha^{(1)} \, g_{E1}(\theta) + 
\delta \beta^{(1)} \, g_{M1}(\theta) \right)  \left( \frac{E}{\hbar c} 
\right)^2 F_1(q).
\end{eqnarray}
The r.h.s of the first line in eqn. (\ref{DeltaM}) contains the ``natural'' 
definition
of the meson exchange corrections since $\delta\alpha$ and
$\delta\beta$ are related to a two-body formfactor.
The definition in the second line is more convenient,
because in this case $\delta\alpha^{(1)}$ and $\delta\beta^{(1)}$ can be
directly 
compared with the  electromagnetic polarizabilities of the free  nucleon.
A calculation of  $\delta\alpha$ and  $\delta\beta$       has been 
carried out by H\"utt  and Milstein \cite{huett96} for nuclear matter.
The result is  $\delta{\alpha}^{(1)}(A = \infty)= - 3.4$
and $\delta{\beta}^{(1)}(A = \infty)= + 
2.4$ (Table 1).  
For finite nuclei the same    authors \cite{huett98}
find the results also listed in Table \ref{MesonCorr}.
%
%
\begin{table}[h]
\caption{\it Meson exchange corrections of the in-medium electromagnetic
polarizabilities versus mass number A in units of
\protect{$10^{-4}\mbox{fm}^3$} 
according to \cite{huett96,huett98}.
For definition see  \mbox{(\ref{DeltaM})}.}
\vspace{3mm}
\begin{center}
\begin{tabular}{|c|c|c|}
\hline
A &\protect{$\delta{\alpha}^{(1)}$}&\protect{$\delta{\beta}^{(1)}$}\\
\hline
$\infty$ & -3.4 & +2.4\\
40 & -2.4 & +2.1\\
16 & -1.3& +1.0 \\
4  & $\approx$ 0 & $\approx$ 0\\
\hline
\end{tabular}
\end{center}
\label{MesonCorr}
\end{table}
We see that the meson exchange corrections of the electromagnetic 
polarizabilities become smaller with decreasing mass number.
This is in line with the expectation that finite-size effects on $\delta 
\alpha$ and $\delta \beta$,  i.e. 
the differences ${\delta \alpha}^{(1)}(A=\infty)-{\delta \alpha}^{(1)}(A)$
and ${\delta \beta}^{(1)}(A=\infty)-{\delta \beta}^{(1)}(A)$,
become larger with smaller mass number A. 
For A = 40 the finite-size corrections do not lead to 
observable effects on the predicted differential cross sections. 
On the other hand, for A = 16 the finite-size 
corrections are large enough to be visible in the data. For A = 4
the meson-exchange corrections $\delta \alpha$ and $\delta \beta$
are comparable with zero. 
This is  in good agreement with the  meson-exchange corrections 
predicted for the  deuteron \cite{levchuk97} which was 
mentioned in the introduction. 
\section{Results and Discussion}
In this section we will discuss the present and previous experimental 
differential cross sections 
obtained for $^{40}$Ca, $^{16}$O and $^4$He together with predictions.
For the calculation of the resonance amplitude $R(E,\theta)$ Lorentzian
representations of the nuclear photoabsorption cross section  are needed.
For $^{16}$O and $^4$He these Lorentzian representations have already 
been published in tabular form in \cite{haeger95} and \cite{fuhrberg95},
respectively.
For $^{40}$Ca the corresponding data are given in Table \ref{CaLorentz}. 
\begin{table}[h]
\caption{\it Parameters and multipolarities of Lorentzian lines used to 
represent the photoabsorption cross section of $^{40}$Ca. QD denotes
the quasi-deuteron cross section.$E_0$: resonance energy; $\sigma_0$: peak
cross section; $\Gamma$: width of the Lorentzian. }
\vspace{3mm}
\begin{center}
\begin{tabular}{|c||c|c|c|}
\hline
Lorentzian&$E_0$ [MeV]&$\sigma_0$ [mb] & $\Gamma$ [MeV]\\
\hline
\hline
E1&19.5&83.5&4.8\\
E1&28.0&7.1&10.5\\
E2&31.0&1.7&10.9\\
QD&50.0&3.1&100.0\\
\hline
\end{tabular}
\end{center}
\label{CaLorentz}
\end{table}
\subsection{$^{40}$Ca}
The four-detector set-up described in detail in \cite{haeger95}
has been used in the present experiment for measuring the differential 
cross sections at the tagged-photon facility of the  stretcher ring MAX 
(Lund, Sweden) at a photon energy of 74 MeV. 
In a second run at the photon energy 58 MeV only three of the detectors 
were used and  positioned at somewhat different angles. The beam time for 
each of the two experiments was 1 week.
The results  of the two experiments are
shown in Figs. 1  and 2 together with predictions which are represented by
curves. Fig. 1$a$   containing present and previous \cite{wright85}
data, shows the energy
dependence of  differential cross sections at $\theta = 45^{\circ}$. 
At this small scattering angle the dependence of the predicted differential
cross section 
on ${\tilde \alpha}_N - {\tilde \beta}_N$
is small so that these data predominantly serve as a test of the 
photoabsorption cross section entering into  $R(E,\theta)$. It is apparent 
that the present and previous data can be interpreted in a consistent way.
For the prediction of the angular distribution of $R(E,\theta)$ also the
strength and location of the isovector quadrupole resonance
(IVGQR) has to be known (see Table \ref{CaLorentz}). 
There are two  investigations of the IVGQR 
using   the (n,$\gamma$) 
\cite{bergqvist84} and ($\gamma$,n) \cite{sims97}
reactions, leading to E2 strengths centred around E$_{\circ}$= 32 MeV
and 31 MeV, respectively.
We have  calculated  differential cross  sections 
assuming a strength 
of one IVGQR  energy-weighted sum rule. 
Since this assumption may possibly be  
uncertain within about 40\% we investigated the 
effects of a variation of the IVGQR strength on the predictions at
$\theta = 135^{\circ}$.
We found that 
for the data interpretation described in the next  paragraph
the precise strength of the IVGQR is not of relevance.
 
Figs. 2$a$ and $b$ show angular distributions of the  differential cross
sections 
obtained in the present experiment together with predictions.
All amplitudes entering into the binding correction 
$S(E,\theta)+R(E,\theta)$ are fixed through procedures
described above so that the only free parameter is the difference
${\tilde \alpha}_N - {\tilde \beta}_N$ with 
${\tilde \alpha}_N + {\tilde \beta}_N$
kept constant. For the three curves
different  effective electromagnetic polarizabilities
(${\tilde \alpha}_{eff}$;${\tilde \beta}_{eff}$) have been 
adopted. These
quantities are defined such that they may be inserted into into eq.
(\ref{NuclAmpl}) instead of $({\tilde \alpha}_N;{\tilde \beta}_N)$ in order
to take care of different tentative choices for the in-medium
electromagnetic polarizabilities including the effects of meson-exchange 
currents and of modifications of the internal structure of the nucleon,
while simultaneously the mesonic seagull-amplitude $S(E,\theta)$ is 
represented through its static approximation ${\tilde S}(E,\theta)$.
The numbers used for the effective polarizabilities are
($i$)\,(11.3;3.7), ($ii$)\,(8.9;5.8) and ($iii$)\,(3.3;11.7) 
corresponding to ($i$)
the free-nucleon values, ($ii$) the same quantities as in ($i$)
but including the meson
exchange 
corrections (see Table \ref{MesonCorr}) and ($iii$) the  
same quantities as in ($i$)  but  including 
the polarizability changes of $\Delta{\alpha}=-8$  
and $\Delta{\beta}=+8$ suggested in \cite{feldman96}.
All three curves are in good agreement with each other and with
the data points at $\theta =45^{\circ}$ but differ to a sizable  extent
from each other at larger angles. The differences  of 
the strength of the IVGQR \cite{bergqvist84,sims97} discussed above and the
different choices for $F_2$(q) 
lead to modifications
of the curves smaller than the difference between curves ($i$, short dashes) 
and ($ii$, solid curves)
and, thus, are irrelevant for our conclusion that 
in-medium modifications as large as
$\Delta{\alpha}=-8$ 
and $\Delta{\beta}=+8$  \cite{feldman96}, corresponding to the upper curve
($iii$, long dashes) are  ruled out by our data. The same conclusion
may be drawn from Fig. 1$b$ where our data for $\theta$ = 135$^{\circ}$
are shown together with those of \cite{wright85}.

Our summary is that the present and earlier \cite{wright85}  obtained data 
for $^{40}$Ca are  
consistent with unmodified  electromagnetic polarizabilities after the 
comparatively small meson-exchange corrections as given by $\delta \alpha$ 
and $\delta \beta$ have been carried out.
This is in agreement with our previous finding where experiments on 
$^{12}$C and $^{16}$O
\cite{ludwig92,haeger95}
led to in-medium electromagnetic polarizabilities which are essentially 
consistent
with the free-nucleon values. 
\subsection{$^{16}$O}
In our present experiment on $^{16}$O carried out at an average
photon energy of 58 MeV the number of counts accumulated 
in one week of  beam time was large enough to provide differential
cross sections with high statistical precision. On the other hand
the high rate per tagger channel discussed in section 2 caused us to 
treat these differential cross sections as relative numbers which had to 
be adjusted to our previous data \cite{haeger95} at small angles by 
applying some common factor. Nevertheless, these new data may serve as a test
of the angular distribution of  differential cross sections
measured in our previous experiment. The results are shown in Figure
3$a$ together with predictions. Apparently, the two sets of experimental data
nicely support each other as far as the angular distribution is concerned.

For the interpretation of the data of Figures 3$a$ and 3$b$ three choices
for the meson exchange corrections $\delta \alpha^{(1)}$ and 
$\delta \beta^{(1)}$ have been adopted.   The lower curves have been
calculated with the meson-exchange corrections set to zero, 
the upper curves by adopting  the nuclear-matter predictions. For the 
curves in the middle the finite-size predictions for A=16 contained in  
Table \ref{MesonCorr} have been used. 
Calculations based on the large in-medium modifications 
$\Delta \alpha = -8$ and $\Delta \beta = +8$ reported in \cite{feldman96}
have not been included here, because this case has been well documented 
in \cite{feldman96} and the inclusion  of a further curve in Figure 3
may  lead to a reduction of clarity.
It is apparent that the curves including the correction
$\delta \alpha^{(1)}(A=16),
\delta \beta^{(1)}(A=16)$ 
provide the optimum fit to the majority of the experimental data,
i.e. if we exclude the two data points for $E_{\gamma} = 75 \, \mbox{MeV}$
at about $\Theta = 60^{\circ}$.   
This finding leads to the conclusion that the predicted 
meson-exchange corrections  are in line with our experiments 
and that there are no indications of additional 
sizable in-medium modification of the
electromagnetic polarizabilities. 
\subsection{$^4$He}
For $^4$He two different series of experiments  have been carried out at 
the MAX accelerator in Lund.  The results of these two series are shown 
in Figures 4 and 5 together with the data of Wells \cite{wells90,wells92}.
The two series of the present experiment are in good agreement with 
each other but appreciably differ from those of \cite{wells90} at large angles.
For the interpretation we closely follow the prescriptions given in
\cite{fuhrberg95} where a careful analysis of the photoabsorption cross 
section of $^4$He has been carried out. As seen in Table \ref{MesonCorr}
the meson-exchange corrections are consistent with zero for $^4$He.
The results of the calculations are given 
by the curves of Figs. 4 and 5. Though the experimental precision of the data 
obtained for $^4$He is inferior to the ones  obtained for $^{40}$Ca and
$^{16}$O it is possible to state that
the data obtained in the present work for $^4$He are consistent with  
unmodified in-medium electromagnetic polarizabilities, in agreement with 
the results we obtained for $^{16}$O and $^{40}$Ca. We have no information
where the discrepancy between our data and the data of the 
Illinois/Saskatoon groups might come from. The only difference between 
the two methods of data taking  we are aware of is 
that the  Illinois/Saskatoon groups
use much higher (factors of $\approx$5) count rates in the tagger channels
than we used in most of our experiments. 
The very apparent advantage of these
higher count rates is that a larger number of data with high statistical
precision is obtained. On the other hand, higher count rates require
larger corrections due to stolen coincidences and pile-up.

\section{Summary and Conclusions}
The present experiment has led to a satisfactory
consistency in our conclusion concerning
the in-medium electromagnetic polarizabilities. This conclusion is that
the proton-neutron averages of in-medium  electromagnetic polarizabilities
${\tilde \alpha}_N$ and ${\tilde \beta}_N$  are the same as the 
corresponding quantities
${\bar \alpha}_N$ and ${\bar \beta}_N$ for the free nucleon within a
precision of the order of $\pm 2.5\times10^{-4}fm^3$, if we do not put
too much weight on some points discussed in the next paragraph.
Furthermore, there are distinct indications that the data obtained for 
$^{40}$Ca and $^{16}$O are in favour of the predicted  \cite{huett98}
meson-exchange corrections $\delta \alpha$ and $\delta \beta$ of the 
electromagnetic polarizabilities.
It was known before
that the sum of electromagnetic polarizabilities 
${\bar \alpha_N}+{\bar \beta_N}$ is not modified through binding.
The essential additional message
of our result is that the difference of electromagnetic polarizabilities
${\bar \alpha_N}-{\bar \beta_N}$ also remains the same. In the dispersion 
theory of nucleon Compton scattering \cite{lvov97} the difference 
${\bar \alpha_N}-{\bar \beta_N}$  is an independent input parameter
which may be related to the exchange of the scalar $\sigma$ meson in the 
t-channel. On the other hand the scalar $\sigma$  is believed to be 
responsible for the largest part of  the binding potential between nucleons
in a complex nucleus. 
This observation suggests that there apparently is a relation between
the quantity ${\tilde \alpha_N}-{\tilde \beta_N}$
measured in the present experiment and the nucleon binding potential. The 
tentative 
conclusion from our experimental results, therefore, may be that the 
r\^ole  the $\sigma$ meson plays for the structure of the nucleon
is not modified when the nucleons are embedded in nuclear matter.

The conclusions drawn in the foregoing section are weakened to some extent
by some
unsolved problems which should be cleared up in further studies:
(i) The free-nucleon polarizabilities of the neutron are not known with the
desirable reliability.
(ii) There is a definite discrepancy between some experimental data measured
by the G\"ottingen/Lund  group and data measured by the 
Illinois/Saskatoon group for $^{16}$O and $^4$He at the large scattering 
angle of $\theta$ = 135$^{\circ}$.
(iii) The  angular distribution of Compton-scattering differential
cross-sections measured by the G\"ottingen/Lund group for $^{12}$C and
$^{16}$O at an energy of 75 MeV deviates from the predictions at
$\theta$ = 60$^{\circ}$ by 20 - 30\% \cite{haeger95}. These deviations 
have not been removed by the recent theoretical studies
of H\"utt and Milstein \cite{huett96,huett98}. 
\section*{Acknowledgement}
The authors are indebted to A.M. Nathan for the permission to make use
of unpublished data on $^4$He.
\newpage
\begin{figure}[t]
 \centering
  \subfigure[Scattering angle $\Theta = $ 
  45$^{\circ}$]{\psfig{figure=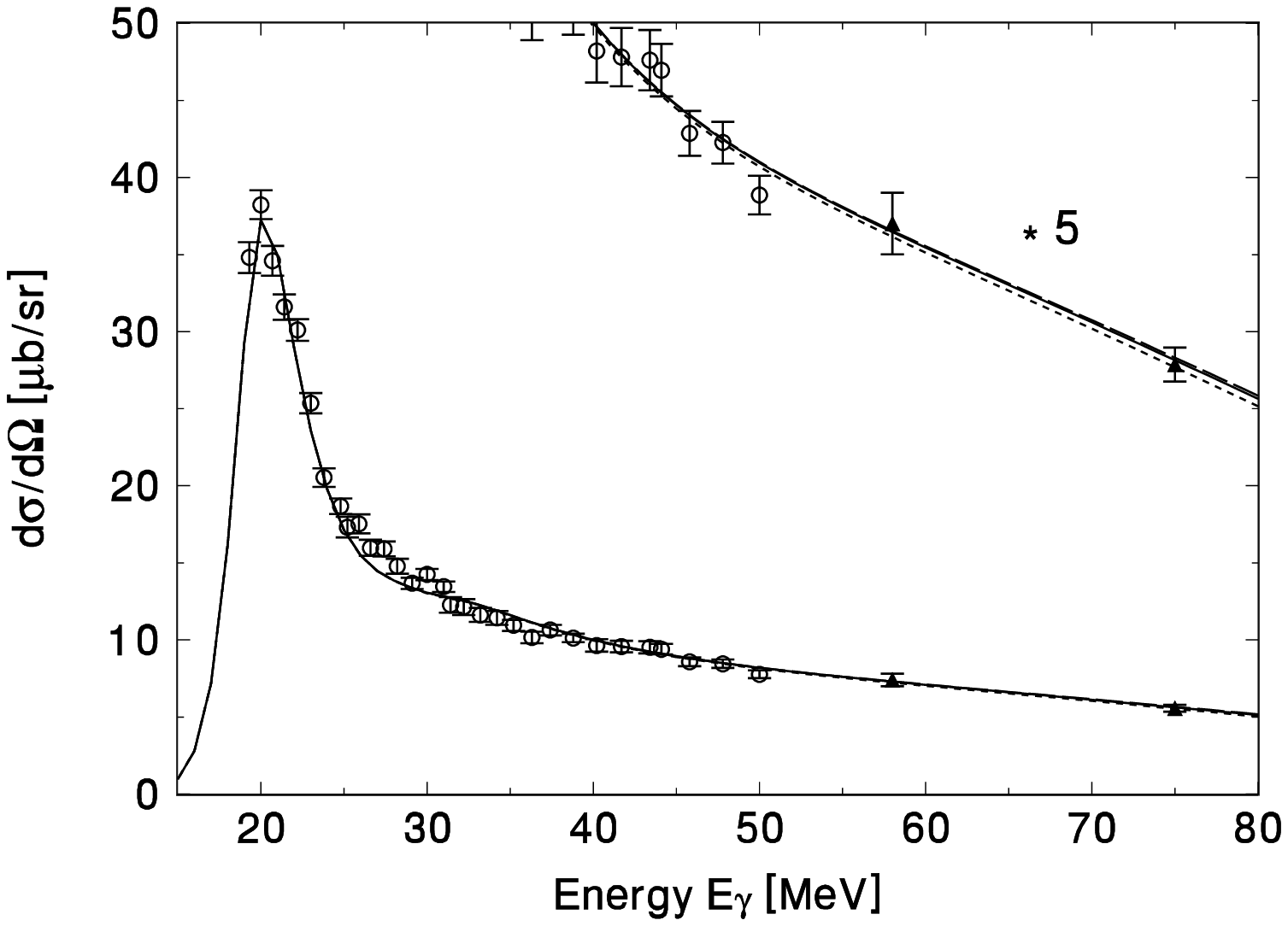,width=11cm}}
  \subfigure[Scattering angle $\Theta = $ 
  135$^{\circ}$]{\psfig{figure=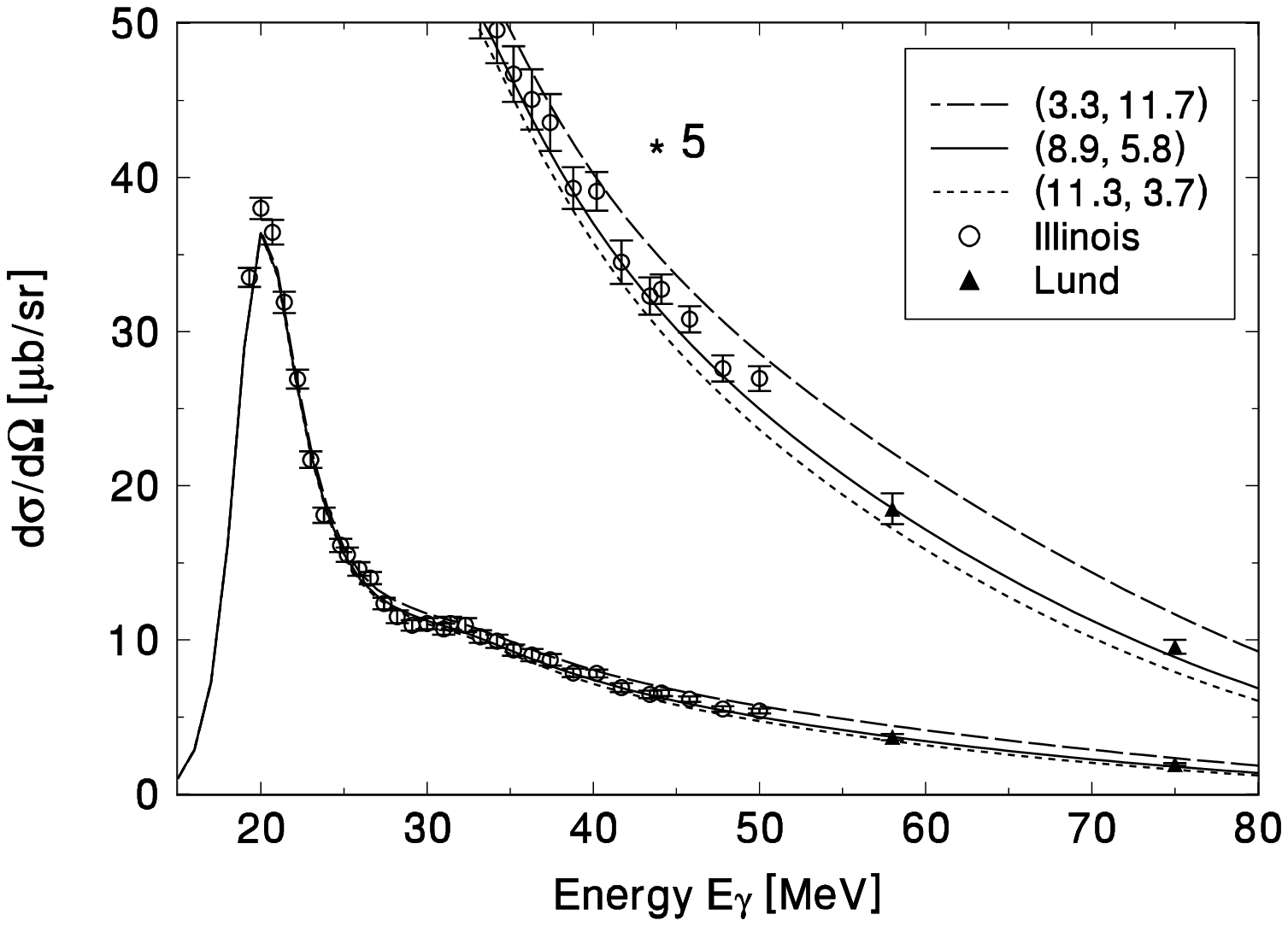,width=11cm}}
\caption{ 
Experimental elastic differential coss sections for $^{40}$Ca
versus photon energy compared with predictions. The data below 50 MeV
are from Wright et al. \cite{wright85} the data above 50 MeV from the 
present work. The curves are calculated for  effective electromagnetic 
polarizabilities ($\tilde{\alpha}_{eff}$;$\tilde{\beta}_{eff}$)
of (3.3;11.7), (8.9;5.8) and (11.3;3.7)  (see insert in Fig. 1$b$). 
The experimental data at 74 MeV are 
shifted in scattering angle in proportion to the predicted differential
cross sections.
}
\end{figure}
\newpage
\begin{figure}[t]
  \centering
  \subfigure[Energy $E_{\gamma} = $ 58MeV]{\psfig{figure=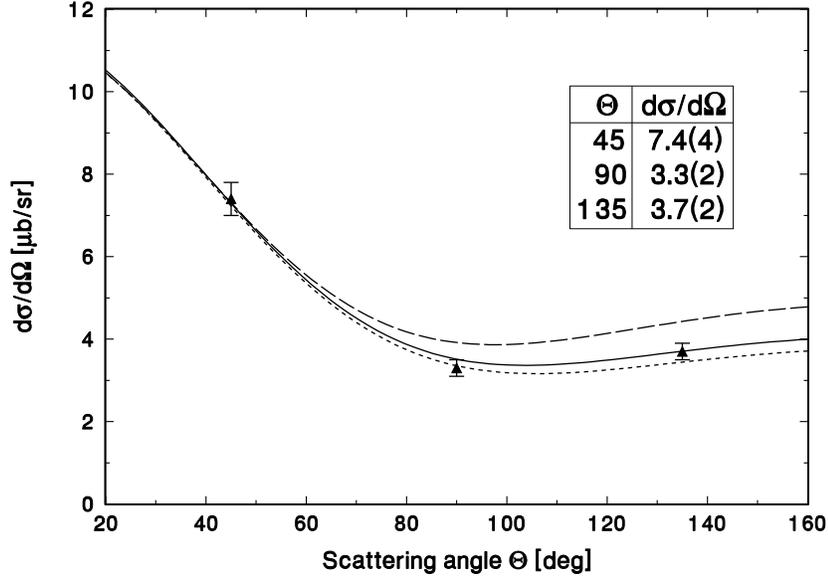,width=12cm}}
  \subfigure[Energy $E_{\gamma} = $ 74MeV]{\psfig{figure=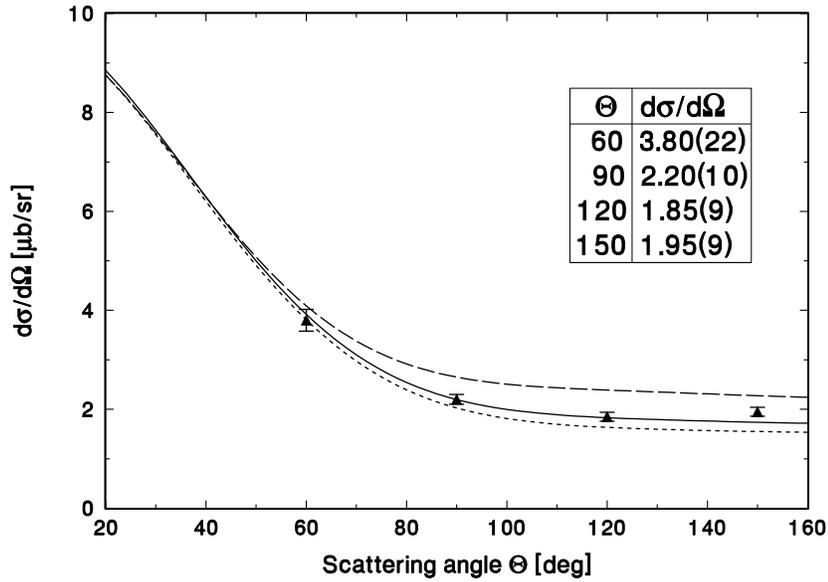,width=12cm}}
\caption{ 
Experimental elastic differential coss sections for $^{40}$Ca
versus scattering angle compared with predictions. The data are from the 
present work. The curves are calculated for the same electromagnetic 
polarizabilities  as in Figure 1 (see insert in Figure 1b). 
The inserts contain the present experimental data in $\mu$b/sr.
}
\end{figure}
\newpage
\begin{figure}[t]
  \centering
 \subfigure[Energy $E_{\gamma} = $ 58MeV]{\psfig{figure=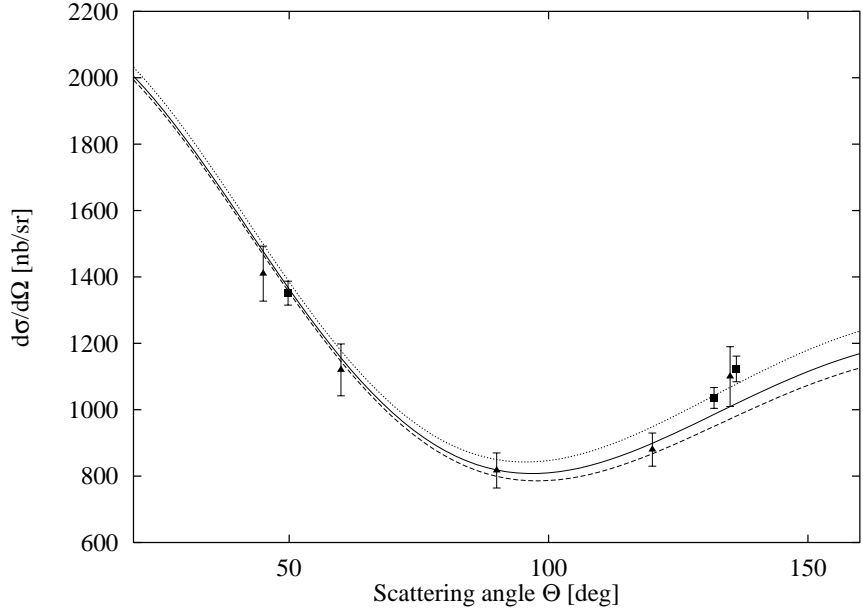,width=12cm,angle=-90}}
  \subfigure[Energy $E_{\gamma} = $ 75MeV]{\psfig{figure=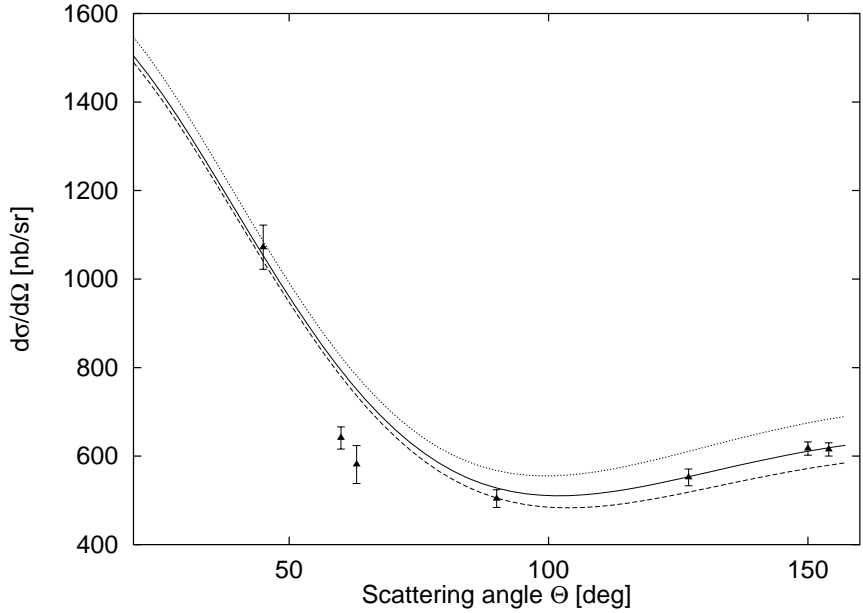,width=12cm,angle=-90}}
\caption{ 
Experimental elastic differential cross sections for $^{16}$O
versus scattering angle compared with predictions. The data are from
our previous work \cite{haeger95} except for the three data points
in Figure 3$a$ depicted by squares which are from the present work. 
The curves are calculated
for different choices of meson exchange corrections 
($\delta \alpha^{(1)}$,$\delta \beta^{(1)}$). Lower curve:
(0,0). Curve in the middle: ($-$1.3,+1.0). Upper curve: ($-$3.4,+2.4).
}
\end{figure}
\newpage
\begin{figure}[t]
  \centering
  \subfigure[Scattering angle $\Theta = $ 45$^{\circ}$]{\psfig{figure=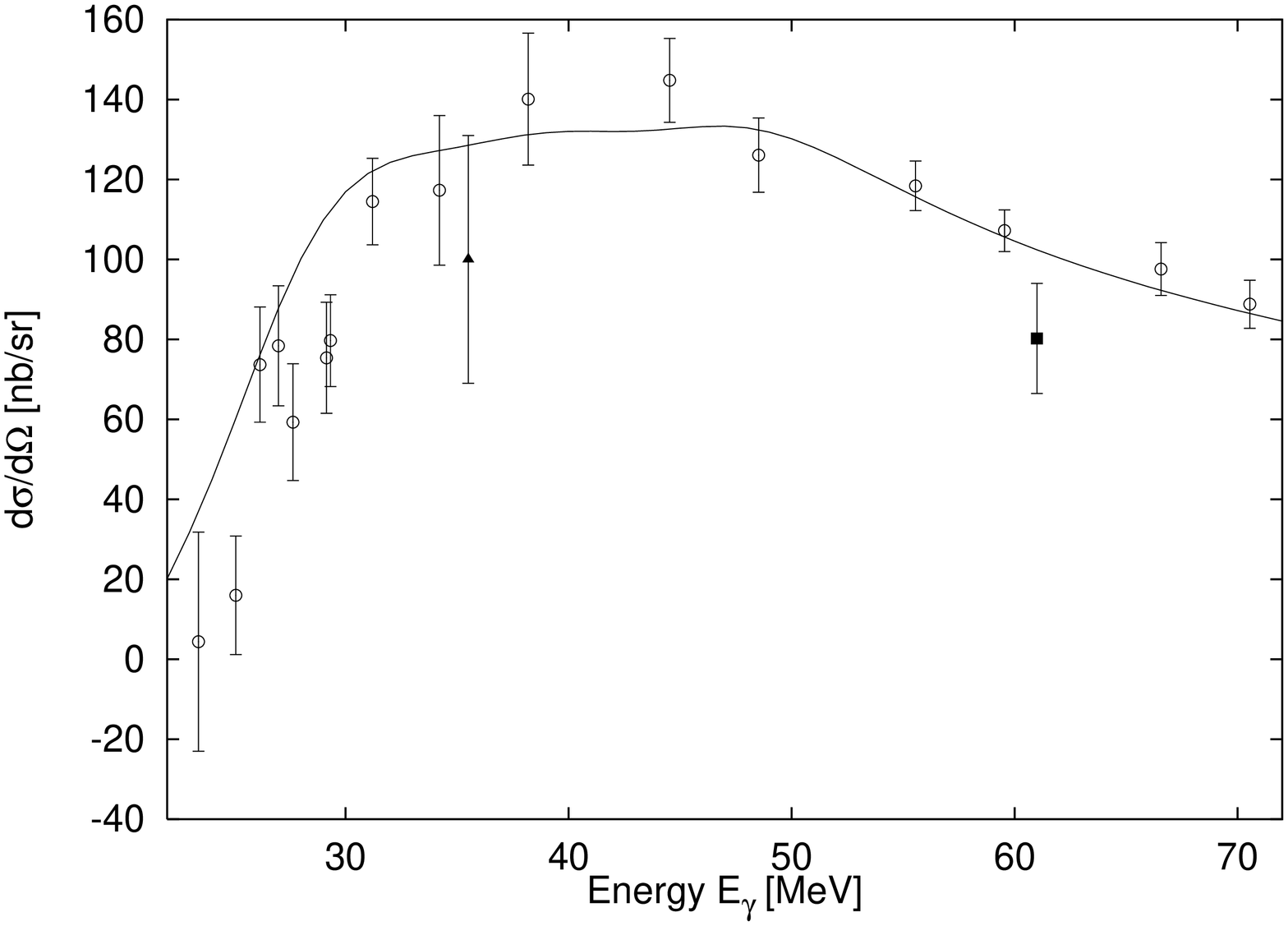,angle=-90,width=12cm}}
  \subfigure[Scattering angle $\Theta = $ 135$^{\circ}$]{\psfig{figure=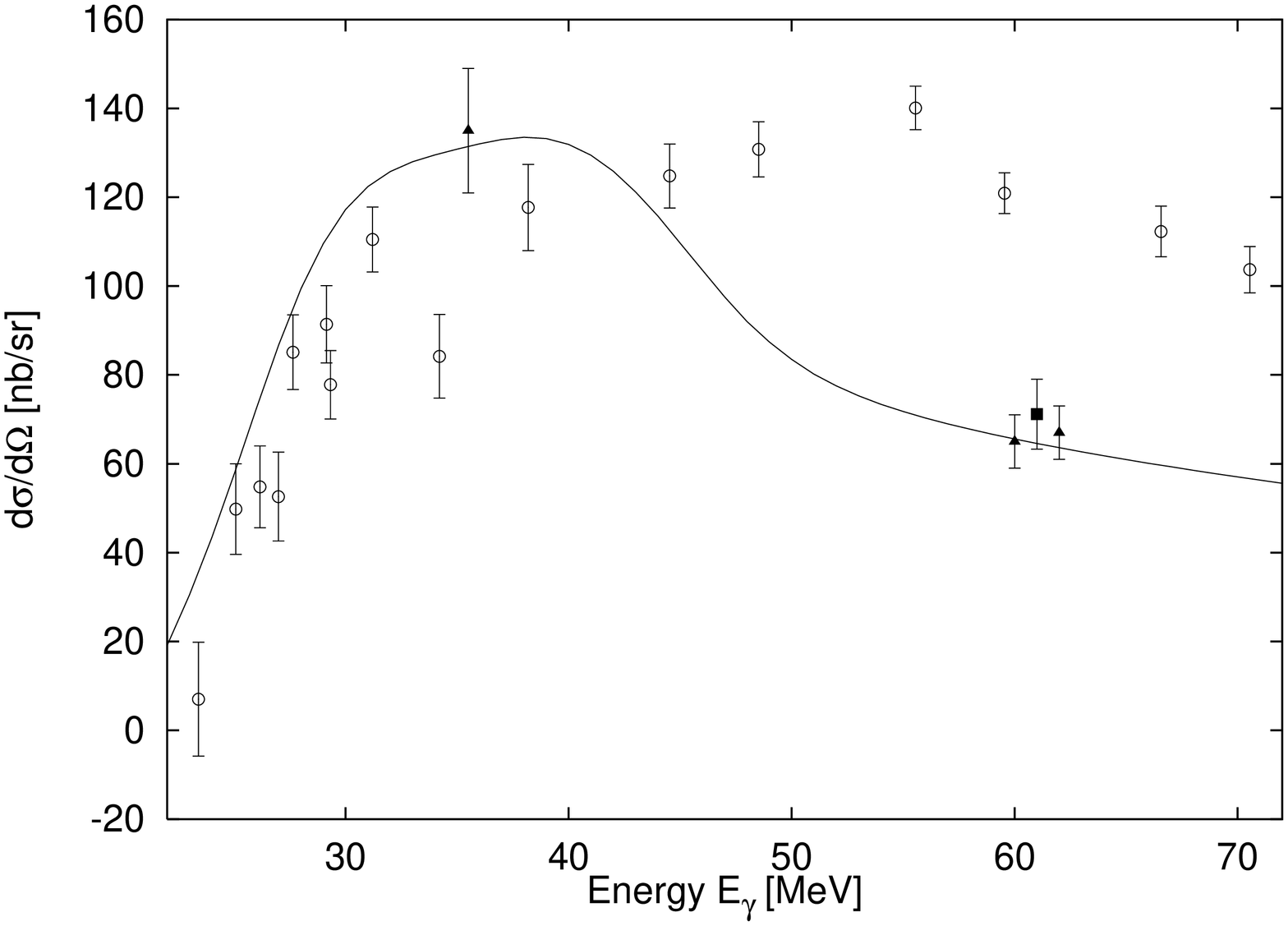,angle=-90,width=12cm}}
\caption{ 
Differential cross sections for Compton scattering by $^4$He versus energy.
The data of the present series of experiments 1 are given by full triangles, 
those of the present series of experiments 2 by full squares, the previous 
data \cite{wells90} by open circles. The curve has been calculated 
according to the prescriptions described in the text.
}
\end{figure}
\newpage
%
\begin{figure}[t]
  \centerline{\psfig{figure=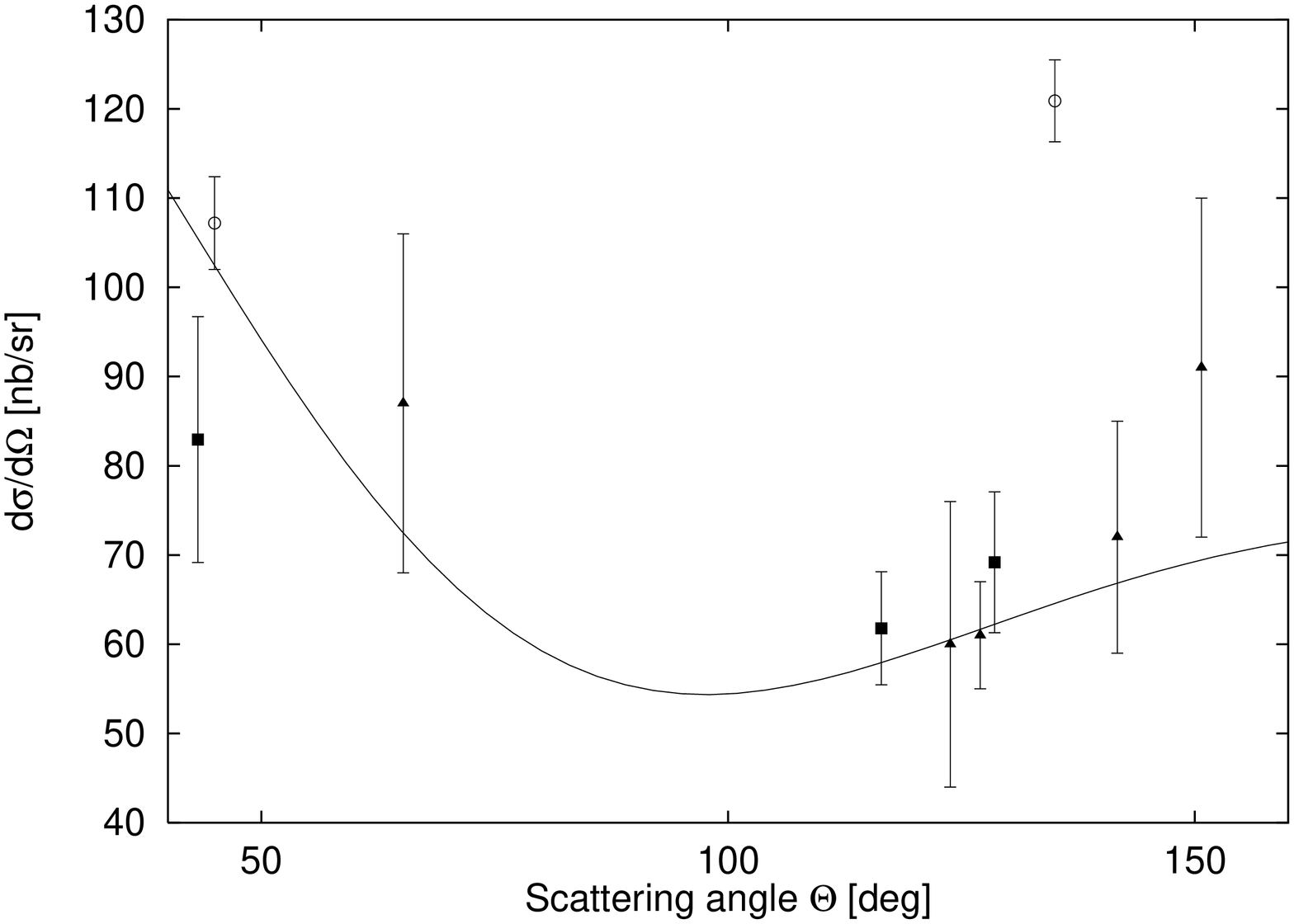,angle=-90,width=12cm}}
  \caption{ 
Differential cross sections for Compton scattering by $^4$He
versus scattering angle at a photon energy of 61 MeV. The data of the 
present series of experiments  1 are given by full triangles, those of the 
present series of experiments 2 by full squares, the previous 
data \cite{wells90} by open circles. The curve has been calculated 
according to the prescriptions described in the text.}
\end{figure}
\end{document}